\documentclass[10pt, twocolumn]{article}
\usepackage[margin=2cm]{geometry}

\pdfoutput=1 

\usepackage[utf8]{inputenc}
\usepackage[english]{babel}

\usepackage{amsthm}   
\usepackage{amsmath}  
\usepackage{array} 
\usepackage{multirow} 
\usepackage{booktabs} 
\usepackage[dvipsnames]{xcolor} 
\usepackage{tikz-cd} 
\usepackage{xurl} 
\usepackage{balance} 
\usepackage{orcidlink}

\usepackage{hyperref}
\hypersetup{
  colorlinks=true,
  citecolor=darkgray,
  linkcolor=darkgray,
  urlcolor=darkgray,
  breaklinks=true,
  bookmarksnumbered=true,
  bookmarksopen=true,
  pdftitle={UTC Time, Formally Verified},
  pdfauthor={
    Ana de Almeida Borges,
    Mireia González Bedmar,
    Juan Conejero Rodríguez,
    Eduardo Hermo Reyes,
    Quim Casals Buñuel,
    Joost J. Joosten
  },
  pdfsubject={},
  pdfcreator={Ana de Almeida Borges},
  pdfkeywords={}
}
\usepackage[numbers]{natbib}   

\newtheorem{theorem}{Theorem}[section]
\newtheorem{remark}[theorem]{Remark}

\newcommand{\CustomForall}[1]{
  \forall #1 \,
}

\usepackage{xifthen}

\newcommand{\codeword}{} 
\DeclareRobustCommand{\codeword}[1]{\normalfont{\texttt{\detokenize{#1}}}}

\newcommand{\urlhash}{\#}
\newcommand{\coqdocurl}[2]{\coqdocbaseurl #1.html\urlhash #2}
\makeatletter
\newcommand{\coqident}{\begingroup\@makeother\#\@coqident}
\newcommand{\@coqident}[3][]{%
  \ifthenelse{\isempty{#2}}%
  {\codeword{#3}}%
  {\ifthenelse{\isempty{#1}}%
  {\href{\coqdocurl{#2}{#3}}{\codeword{#3}}}%
  {\href{\coqdocurl{#2}{#3}}{\codeword{#1}}}}%
\endgroup}
\newcommand{\coqfile}[2]{%
  \ifthenelse{\isempty{#1}}%
    {\href{\coqdocbaseurl #2.html}{\codeword{#2.v}}}%
    {\href{\coqdocbaseurl #1.#2.html}{\codeword{#2.v}}}}

\makeatother

\newcommand{\coqdocUTC}{https://formalv.gitlab.io/-/formalv/-/jobs/4677927207/artifacts/public/coqdoc/formalv.} 
\newcommand{\coqdocbaseurl}{\coqdocUTC}
\newcommand{\coqstdliblink}{https://coq.inria.fr/distrib/V8.17.1/stdlib} 
\newcommand{\mathcomplink}{https://math-comp.github.io/htmldoc_1_17_0} 

\newcommand{\Uint}{%
  \href{\coqstdliblink/Coq.Numbers.Cyclic.Int63.Uint63.html}{%
    \codeword{Uint63}%
  }%
}
\newcommand{\Sint}{%
  \href{\coqstdliblink/Coq.Numbers.Cyclic.Int63.Sint63.html}{%
    \codeword{Sint63}%
  }%
}

\newcommand{\ExtrOcamlBasic}{%
  \href{\coqstdliblink/Coq.extraction.ExtrOcamlBasic.html}{%
    \codeword{ExtrOcamlBasic}%
  }%
}
\newcommand{\ExtrOcamlInt}{%
  \href{\coqstdliblink/Coq.extraction.ExtrOCamlInt63.html}{%
    \codeword{ExtrOCamlInt63}%
  }%
}

\newcommand{\ssrnat}{%
  \href{\mathcomplink/mathcomp.ssreflect.ssrnat.html}{%
    \codeword{ssrnat}%
  }%
}
\newcommand{\ssrint}{%
  \href{\mathcomplink/mathcomp.algebra.ssrint.html}{%
    \codeword{ssrint}%
  }%
}

\newcommand{\fintype}{%
  \href{\mathcomplink/mathcomp.ssreflect.fintype.html}{%
    \codeword{fintype}%
  }%
}

\newcommand{\nat}{%
  \href{\coqstdliblink/Coq.Init.Datatypes.html\urlhash nat}{%
    \codeword{nat}%
  }%
}

\newcommand{\bool}{%
  \href{\coqstdliblink/Coq.Init.Datatypes.html\urlhash bool}{%
    \codeword{bool}%
  }%
}

\newcommand{\Cal}[1]{\coqident{time.calendar}{#1}}
\newcommand{\Hin}[1]{\coqident{time.Hinnant}{#1}}

\newcommand{\calendar}[1]{\codeword{calendar}.\allowbreak \Cal{#1}}
\newcommand{\Hinnant}[1]{\codeword{Hinnant}.\allowbreak \Hin{#1}}

\newcommand{\Metacoq}{%
  \href{https://metacoq.github.io/}{%
    \textsf{MetaCoq}%
  }%
}

\title{UTC Time, Formally Verified}
\author{
  Ana de Almeida Borges\,\orcidlink{0000-0001-5152-198X}
  \and Mireia González Bedmar\,\orcidlink{0000-0002-9188-917X}
  \and Juan Conejero Rodríguez\,\orcidlink{0000-0002-7801-9532}
  \and Eduardo Hermo Reyes\,\orcidlink{0000-0002-8982-6030}
  \and Joaquim Casals Buñuel\,\orcidlink{0009-0005-6318-3726}
  \and Joost J. Joosten\,\orcidlink{0000-0001-9590-5045}
}
\date{University of Barcelona and Formal Vindications S.L.}

\begin{document}

\maketitle

\begin{abstract}
FV Time is a small-scale verification project developed in the Coq proof assistant using the Mathematical Components libraries. It is a library for managing conversions between time formats (UTC and timestamps), as well as commonly used functions for time arithmetic. As a library for time conversions, its novelty is the implementation of leap seconds, which are part of the UTC standard but usually not implemented in existing libraries. Since the verified functions of FV Time are reasonably simple yet non-trivial, it nicely illustrates our methodology for verifying software with Coq.

In this paper we present a description of the project, emphasizing the main problems faced while developing the library, as well as some general-purpose solutions that were produced as by-products and may be used in other verification projects. These include a refinement package between proof-oriented MathComp numbers and computation-oriented primitive numbers from the Coq standard library, as well as a set of tactics to automatically prove certain decidable statements over finite ranges through brute-force computation.
\end{abstract}


\section{Introduction}\label{sec1}
\label{sec:intro}

\begin{figure*}[!ht]
  \centering
  \begin{tikzcd}[row sep=3ex, column sep=tiny]
    \coqfile{time}{calendar} && \\
    && \\
    \coqfile{time}{Hinnant} \arrow[uu, "\text{implements}"] && \coqfile{time}{formalTime} \arrow[ll, "\text{uses}", dotted] \\
    && \\
    \coqfile{time}{HinnantR} \arrow[uu, "\text{refines}"] && \coqfile{time}{formalTimeR} \arrow[uu, "\text{refines}" swap] \arrow[ll, "\text{uses}", dotted] \\
    && \\
    & \coqfile{time}{fvtm_extraction} \arrow[ruu, "\text{refines}" swap] \arrow[luu, "\text{refines}"] & \\
    && \\
    & \coqfile{time}{fvtm_extraction_correct} \arrow[uu, "\text{proves correct}"] &
  \end{tikzcd}
  \caption{The file structure of FV Time. There is an extra file, \codeword{doe_of_yoeK.v}, that only contains auxiliary proofs and definitions and doesn't appear in the diagram.}
  \label{fig:structure}
\end{figure*}
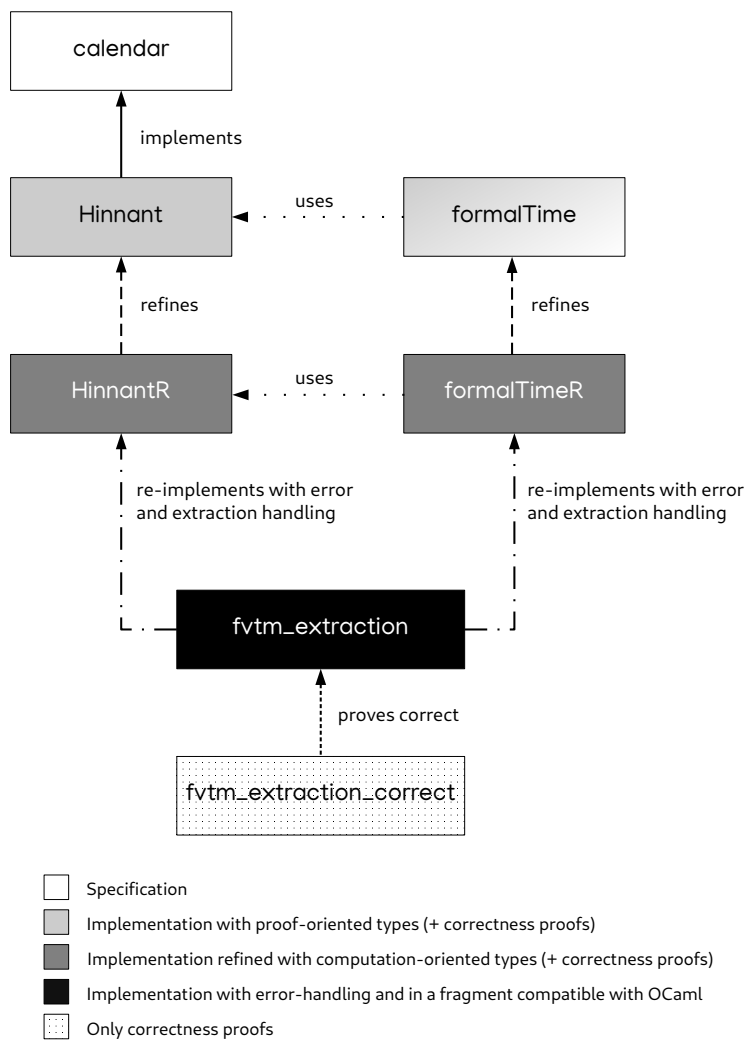

Coordinated Universal Time (UTC) \cite{UTCdef} is the current world standard for keeping time. Although it uses atomic time, it is designed to stay close to solar time, and as such it includes leap seconds. The number of seconds in a minute can be either 59 (if there is a negative leap second), 60 (the regular case), or 61 (if there is a positive leap second). The need for a new leap second is somewhat unpredictable, so the International Earth Rotation and Reference Systems Service announces whether there will be one about six months in advance. The convention is to have at most two leap seconds per year, as the final second of the last day of a month, preferably June or December. As of 2023, there have been 27 positive leap seconds and no negative ones \cite{LeapSeconds}, although a recent resolution \cite{NoMoreLeapSeconds} aims to eliminate future leap seconds, prompted by the various issues and inconveniences they lead to \cite{Levine2023}.

The vast majority of software uses Unix time \cite{microsoftdatetime, androiddate}, which is an implementation of UTC without leap seconds \cite{Unix_spec}. This is fine for many use cases, and understandable given the unpredictability of UTC for future moments. However, it conflicts with legal regulations that explicitly require UTC. It may seem like 27 seconds are not enough to meaningfully change anything, but in fact even 27 seconds can make a difference in real world legal applications, as well as in critical systems. In particular, in the context of software that interprets and evaluates the log information for driving time in the road transport sector according to Regulation (EU) 2016/799 \cite{regulation799}, the algorithm that translates from second-resolution data to minute-resolution data as required by the Regulation can give opposite results depending on whether UTC or Unix is used -- meaning that there exists a possible data file that gets interpreted as 100\% of driving time in UTC and as 0\% of driving time in Unix \cite{TrafficFine_SpecialIssue}.

It was in this context that FV Time \cite{FormalV1.2.0} was developed. It is a by-product of the collaboration between the University of Barcelona and Formal Vindications S.L., whose main goal is the development of large-scale formally verified software with applications in several critical sectors. FV Time is a small-scale verification project developed in Coq and relying in the MathComp library. It includes conversions between time, represented as a 7-tuple of year, month, day, hour, minute, second, and proof of existence in the chosen paradigm (to avoid ill-formed tuples such as the ones including February 30), and timestamps, represented as the number of seconds since a chosen epoch (for example, year 0, or year 1970, which is the Unix epoch). We also define dates and datestamps, which are the respective concepts without information on the hour, minute, and second. Functions for time and duration arithmetic are provided as well. Leap seconds are tracked via a modifiable parameter, which can be empty for Unix time or updated as appropriate to keep pace with UTC.

This paper reports on the development of FV Time from high-level specifications to executable code integrated with other software. It can serve as a roadmap for other similar verification projects.
Section \ref{sec:library} describes the main functions that were verified and our methodology to structure the project, with an emphasis on the issues and solutions that can be of general interest.

We also describe two general purpose Coq libraries: FV Prim63 to MathComp (Section~\ref{sec:utils}) and FV Check Range (Section~\ref{sec:plungin}). These libraries were developed as aids to FV Time, but they would be useful in other contexts as well.
FV Prim63 to MathComp is a collection of results linking Coq primitive integers to MathComp natural numbers and integers, which are needed when refining primitive integers to those MathComp types.
FV Check Range is a small set of automation tactics that solve (provable) goals of the form ``for every primitive integer $x$ in the range $a \leq x < b$, we have $f(x)$'', where $f$ is a boolean function and $a$ and $b$ are fixed primitive integers. Although these kinds of goals can sometimes be automatically solved using preexisting tactics, our approach simply tests every value of $x$ between $a$ and $b$, which works regardless of the boolean function $f$ and is quite fast.

We briefly outline our method of obtaining clean extracted code in Section~\ref{sec:extraction}. The resulting OCaml code was bundled with a command-line interface called FVTM (Section~\ref{sec:fvtm}), which can be used by other applications. This makes our time translations available outside the relatively small world of Coq and OCaml.

Finally, Section \ref{sec:relatedwork} gives an overview of the existing related work, while Section \ref{sec:concl} lists the contributions and conclusions of this project.

\section{FV Time}
\label{sec:library}

\subsection{File Structure}

The main goal of FV Time is to provide verified functions translating between UTC times (with leap seconds) and timestamps. We describe the file structure of the library in Figure~\ref{fig:structure}.

As we see throughout this section, the \coqfile{time}{calendar} file describes the main datatypes, such as what it means to be a UTC time.%
\footnote{Throughout the library and paper, ``time'' refers to our specification of a UTC point expressed as a date-time, i.e., a 7-tuple of year, month, day, hour, minute, second, and proof of validity. We currently take the second to be the smallest unit, although a finer-grained resolution could be implemented as well. This is described in more detail in Section~\ref{sec:datatypes}.}
It also specifies the expected behavior of the translating functions in an intuitive way. We provide a second file, \coqfile{time}{Hinnant}, with alternative implementations of these translating functions. The implementations of the datestamp algorithm and its inverse are inspired by the ones described by Howard Hinnant \cite{Hinnant}, hence the name of the file.

FV Time also provides functions to perform basic arithmetic on UTC times, such as adding a certain number of hours to a given time, in \coqfile{time}{formalTime}.

Since these algorithms are meant to be extracted from Coq to OCaml for efficient execution, we provide a type refinement for each in the \coqfile{time}{HinnantR} and \coqfile{time}{formalTimeR} files. In other words, there are two versions of each algorithm: one based on proof-friendly datatypes, and one based on extraction and computation-friendly ones. These two versions are proven equivalent under some assumptions.

Finally, the two extraction files \coqfile{time}{fvtm_extraction} and \coqfile{time}{fvtm_extraction_correct} follow the extraction method explained in Section~\ref{sec:extraction}.

\subsection{Main Data Types}
\label{sec:datatypes}

The central data type in FV Time is a representation of moments in time in UTC,\footnote{We further assume that the time occurs before the end of year 9999, for reasons explained in Section~\ref{sec:specifications}.} which we call \coqident{time.calendar}{time}. Under the hood it is simply a 6-tuple of natural numbers representing a given year, month, day, hour, minute, and second, together with a proof that the tuple in question forms an existing time. What counts as an existing time depends on the parametrized list of leap seconds.

For convenience and modularity's sake, we define three other relevant types: \coqident{time.calendar}{date}, \coqident{time.calendar}{rawDate}, and \coqident{time.calendar}{rawTime}. A \coqident{time.calendar}{date} is the part of the \coqident{time.calendar}{time} with only the year, month, and day, together with a proof that it exists in UTC. The \codeword{raw} types are simply the tuples without the proofs. Thus, January 32nd 2000 could be represented as a \coqident{time.calendar}{rawDate} but not a \coqident{time.calendar}{date}.

We encode the list of leap seconds as a parameter that can be updated each time a new leap second is announced. The list is actually a list of pairs, where each pair has a date (indicating that a leap second occurs on that date) and a boolean value (where \codeword{false} means that it's a positive leap second and \codeword{true} that it's a negative one). Since we treat the list as a parameter with unknown contents, it can be instantiated in any way as long as it satisfies the hypotheses we use throughout the theorems: the list must be sorted with respect to the strict order of dates (in particular it doesn't include repeated dates), and all the dates in it must be valid. In particular, FV Time can be used to compute Unix time conversions by using an empty list of leap seconds.

The \codeword{raw} types are used in the implementations of every function that operates on dates, such as \coqident{time.Hinnant}{datestamp}. It is then possible to compute the datestamp of January 32nd 2000, but we do not wish to prove any facts about the datestamps of such ill-formed dates. For that reason, we use the valid (non-\codeword{raw}) versions in the specifications and theorem statements. There is then a disconnect between the specification and the implementation, since they refer to different types. This is easily solved using coercions, i.e., automatically inserted translations between one type and another.

\begin{figure}[ht]
\centering
\begin{tikzcd}
  & & \coqident{time.calendar}{time} \arrow[rrdd, "\coqident{time.calendar}{rawTime_of_time}" description] \arrow[lldd, "\coqident{time.calendar}{date_of_time}" description] & & \\
  & & & & \\
  \coqident{time.calendar}{date} \arrow[rrdd, "\coqident{time.calendar}{rawDate_of_date}" description] & & & & \coqident{time.calendar}{rawTime} \arrow[lldd, "\coqident{time.calendar}{rawDate_of_rawTime}" description] \\
  & & & & \\
  & & \coqident{time.calendar}{rawDate} & &
\end{tikzcd}
\caption{A representation of the four main data types in FV Time and the coercions between them. Each arrow from \codeword{X} to \codeword{Y} represents the coercion \codeword{Y_of_X}.}
\label{fig:diamond}
\end{figure}
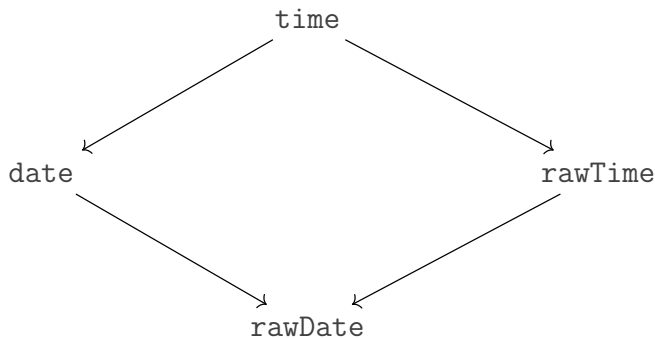

We use a number of coercions in our development, mostly between types and their subtypes, as described in Figure~\ref{fig:diamond}.
We have a very small type hierarchy. Formalizations of, say, mathematical algebra or large libraries such as MathComp include rich hierarchies~\cite{Sakaguchi2020}, and there are existing tools to implement and maintain such large hierarchies such as Hierarchy Builder~\cite{HierarchyBuilder}.

Still, even with a small hierarchy we do run into some issues.
For example, looking at Figure~\ref{fig:diamond}, we can see that there are two possible paths from a \coqident{time.calendar}{time} to a \coqident{time.calendar}{rawDate}. It happens that these paths are definitionally equivalent, and so in some contexts it is irrelevant which one is chosen. However, sometimes the information that the \coqident{time.calendar}{date} part of the \coqident{time.calendar}{time} is valid is crucial, and so the path that goes through the \coqident{time.calendar}{date} must be picked over the other one. In particular, the following unification problem sometimes arises:
\begin{align}
  \label{eq:unification}
  \begin{split}
  &\coqident{calendar}{rawDate_of_date} \ ?d
  = \\ & \hspace{5mm} \coqident{calendar}{rawDate_of_rawTime} \ (\coqident{calendar}{rawTime_of_time} \ t)
  \end{split}
\end{align}

In words, given a \coqident{time.calendar}{time} $t$, a \coqident{time.calendar}{date} $?d$ must be found such that its \coqident{calendar.time}{rawDate} corresponds to the \coqident{time.calendar}{rawDate} of the \coqident{time.calendar}{rawTime} of the \coqident{time.calendar}{time}.
The diamond represented in Figure~\ref{fig:diamond} trivially commutes, so we define the canonical coercion \coqident{time.calendar}{date_of_time} as the path to solve \eqref{eq:unification} with $?d := \coqident{time.calendar}{date_of_time} \ t$.

\subsection{Main Functions}
\label{sec:main_functions}

The backbone of FV Time is the translations between times and timestamps, which themselves depend on translations between dates and datestamps. In this section we focus on the specification and implementation of these four functions, as well as the proofs that they coincide on well-formed inputs. These terms are listed in Tables~\ref{tab:main_functions:types} and \ref{tab:main_functions:correctness}.

\begin{table*}[t]
\centering
\caption{Names and types of the main functions in FV Time.}
\label{tab:main_functions:types}

\begin{tabular}{lll}
\toprule
 & Specification & Implementation \\
 & \coqfile{time}{calendar} & \coqfile{time}{Hinnant} \\ \midrule
  \multirow{2}{*}{date}
    & $\Cal{datestamp} : \Cal{date} \to \codeword{'I_}\Cal{max_datestamp}\codeword{.+1}$
    & $\Hin{datestamp} : \Cal{rawDate} \to \nat{}$
    \\
    & $\Cal{from_datestamp} : \nat{} \to \Cal{date}$
    & $\Hin{from_datestamp} : \nat{} \to \Cal{rawDate}$
    \\
    \midrule
\multirow{2}{*}{time}
    & \begin{tabular}{@{}l@{}}
        $\Cal{timestamp} : \forall(ls : \Cal{leapSeconds}),$\\
        $\ \ \ \ \Cal{time}\ ls \to \codeword{'I_}(\Cal{max_timestamp}\ ls)\codeword{.+1}$
      \end{tabular}
    & $\Hin{timestamp} : \Cal{leapSeconds} \to \Cal{rawTime} \to \nat{}$
    \\
    & \begin{tabular}{@{}l@{}}
        $\Cal{from_timestamp} : \forall(ls : \Cal{leapSeconds}),$\\
        $\ \ \ \ \nat{} \to \Cal{time}\ ls$
      \end{tabular}
    & $\Hin{from_timestamp} : \Cal{leapSeconds} \to \nat{} \to \Cal{rawTime}$
    \\
 \bottomrule
\end{tabular}
\end{table*}

The relevant files are \coqfile{time}{calendar} and \coqfile{time}{Hinnant}. The former includes the basic definitions of dates and times, as well as all the specifications. The latter includes the efficient implementations and the correctness proofs of the main functions. Note that the specification and implementation of a given function have the same name, so we use the name of the file to clarify which we mean at any given time. Similarly, some lemma names coincide and are thus clarified as well.

\subsubsection{Specifications}
\label{sec:specifications}

The specifications of the main functions were primarily chosen to be intuitive. Thus, the \calendar{datestamp} of a \coqident{time.calendar}{date} $d$ is the size of the set of dates strictly smaller than $d$, and similarly for \calendar{timestamp}, where the order relations on dates and times are defined as expected. These definitions use the notion of cardinality of a finite set, which is defined in MathComp's \fintype{} library \cite{mahboubi2013}. In order to benefit from this library's theory, our types for valid dates and times needed bounds so that they could be declared as a \codeword{finType}. While a minimum was already imposed by our definition using natural numbers (a year before 0 cannot be expressed), we arbitrarily set the maximum date as December 31st, 9999. Different end years could be substituted, although if they were large enough there might be problems with overflow during the refinement process (see Section~\ref{sec:refinements}).

The \calendar{datestamp} function goes from \coqident{time.calendar}{date}, the type of valid dates, to \codeword{'I_max_datestamp.+1}, the type of natural numbers smaller than $\coqident{time.calendar}{max_datestamp} + 1$. Hence, on the specification side, the type of the function already ensures that the argument and the result are in the expected range.

To specify the inverse, we start by defining a notion of \coqident{time.calendar}{next_date}, which goes as expected: add 1 to the day component if doing so results in a valid date; otherwise set the day to 1 and add 1 to the month component if doing so results in a valid date; otherwise set the day to 1, the month to January and add 1 to the year unless the year was 9999 (the maximum year), in which case set the year to 1. Its only particularity is that the successor of the maximum date is the minimum date. This cyclic behavior was chosen as a convenient, simple way to maintain the invariant that the successor of a valid date is always a valid date. We also define \coqident{time.calendar}{next_time} to be cyclic, and we underline that this definition depends on the parametrized list of leap seconds, as do essentially all the functions related to time.

Given these notions of \coqident{time.calendar}{next_date} and \coqident{time.calendar}{next_time}, we define the \calendar{from_datestamp} of a number $n$ as the $n$th iteration of \coqident{time.calendar}{next_date} after the minimum date, and similarly for \calendar{from_timestamp}. The spirit of \calendar{from_datestamp} is to describe the counting process one would perform on a real calendar: it is defined as if, given a datestamp $n$, one counted the days one by one from the epoch up to $n$, and the last counted day is the result.

\begin{table*}[t]
\centering
\caption{Theorems stating that the implementations of the main functions meet the specifications.}
\label{tab:main_functions:correctness}

\begin{tabular}{ll}
\toprule
 & Correctness \\
 & \coqfile{time}{Hinnant} \\ \midrule
  \multirow{3}{*}{date}
    & \begin{tabular}{@{}l@{}} 
        $\Hin{datestampE} : \forall(d : \Cal{date}),$\\
        $\ \ \ \ \Hinnant{datestamp}\ d = \calendar{datestamp}\ d$
      \end{tabular}
    \\
    & \begin{tabular}{@{}l@{}}
        $\Hin{from_datestampE} : \forall(n : \nat{}), n \leq \Cal{max_datestamp} \to$\\
        $\ \ \ \ \Hinnant{from_datestamp}\ n = \calendar{from_datestamp}\ n$
      \end{tabular}
    \\
    \midrule
\multirow{6}{*}{time}
    & \begin{tabular}{@{}l@{}}
        $\Hin{timestampE} : \forall(ls : \Cal{leapSeconds})(t : \Cal{time}\ ls),$\\
        $\ \ \ \ \codeword{sorted Order.lt (unzip1}\ ls\codeword{)} \to$\\
        $\ \ \ \ \codeword{all}\ \Cal{valid_date}\ \codeword{(unzip1}\ ls\codeword{)} \to$\\
        $\ \ \ \ \Hinnant{timestamp}\ ls\ t = \codeword{@}\calendar{timestamp}\ ls\ t$
      \end{tabular}
    \\
    & \begin{tabular}{@{}l@{}}
        $\Hin{from_timestampE} : \forall(ls : \Cal{leapSeconds})(n : \nat{}),$\\
        $\ \ \ \ \codeword{sorted Order.lt (unzip1}\ ls\codeword{)} \to$\\
        $\ \ \ \ \codeword{all}\ \Cal{valid_date}\ \codeword{(unzip1}\ ls\codeword{)} \to$\\
        $\ \ \ \ n \leq \Cal{max_timestamp}\ ls \to$\\
        $\ \ \ \ \Hinnant{from_timestamp}\ ls\ n = \calendar{from_timestamp}\ ls\ n$
      \end{tabular}

    \\
 \bottomrule
\end{tabular}
\end{table*}

\subsubsection{Implementations}
\label{sec:implementations}

\paragraph{\Hinnant{datestamp}}

If every month had the same number of days and every year the same number of months, computing the datestamp of a given date would be as straightforward as multiplying each date component by its corresponding number of days and adding everything. Even with different lengths for different months it would not be particularly complicated, but the existence of leap years means that some more care must be taken. In order to have the algorithm be as simple as possible, we internally use shifted years that start on the first day of March and end on the last day of February, as inspired by Hinnant~\cite{Hinnant}. Thus, the leap day, if it exists, is the last day of the shifted year and doesn't influence the calculation of the datestamp of any day in that year other than itself. Since the rule for leap years in the Gregorian calendar repeats itself over periods of 400 years, we also compute the era (period of 400 years) corresponding to the date.

The only other main part of \Hinnant{datestamp} is calculating how many days there are between the start of a (shifted) year and the first day of each (shifted) month. This can be represented as a table that assigns the appropriate value to each (shifted) month -- for March (month 0) it is 0, for April (month 1) it is 31, for May (month 2) it is 61, and so on. However, it turns out that the linear equation $(153\cdot m +2)/5$, where $m$ is the ordinal of the (shifted) month, interpolates this table, so we use that instead of storing the table in memory.

\paragraph{\Hinnant{from_datestamp}}

As in the previous case, we divide years in 400 year eras and shift everything so that years start in March. With this framing, the algorithm is more obviously the inverse of \Hinnant{datestamp} (a fact that we rely on for the correctness proofs).

Given the number of days since the beginning of the era, finding the year must take into consideration leap years, and finding the month must take into consideration the varying number of days in each month. For the latter we use a linear interpolant of the table matching days in a year to months instead of using the table directly.

\paragraph{\Hinnant{timestamp}}

This is a natural extension to time of \Hinnant{datestamp}. In the absence of leap seconds, we could simply add the product of each time component by the amount of seconds in that component (60 seconds per minute, and so on). With leap seconds, we need to additionally calculate the offset generated by them, i.e., the number of extra (positive or negative) seconds that must be added to the leap second-less timestamp, and add it accordingly.
Such an offset is relatively easy to calculate for a given date $d$, for we can simply count the number of dates in our list of leap seconds that happened prior to $d$, and then check whether they were positive or negative to obtain a final offset. This is accomplished by the \coqident{time.Hinnant}{offset_rd} function.

Since the leap second offset can \emph{a priori} be negative (even though there hasn't been a single negative leap second as of 2023), we first calculate the timestamp over the integers and then take its absolute value. This works because even before taking the absolute value we know that the timestamp is positive due to our restrictions on leap seconds: we allow at most one leap second per day (an unimportant restriction, since the international convention allows at most two leap seconds per year). Since there are less days than seconds in any given amount of time, it is not possible to have enough negative leap seconds to obtain a negative timestamp.

\paragraph{\Hinnant{from_timestamp}}

Once we know how to calculate the date corresponding to some datestamp, calculating the time corresponding to some timestamp (i.e., to some number $n$ of seconds since the epoch) is straightforward in the absence of leap seconds. Thus, we first subtract the relevant offset from $n$ and then proceed as if there were no leap seconds.

Obtaining the offsets for this function is slightly more complicated than it was for \Hinnant{timestamp}, since our list of leap seconds is a list of dates and not of timestamps. Thus, the offset calculator for \Hinnant{from_timestamp}, called \coqident{time.Hinnant}{offset_ts}, first computes the \Hinnant{timestamp} of the final second of each date in our list of leap seconds (leap seconds are always the final second of each day by international convention), and then proceeds similarly to the offset computation for \Hinnant{timestamp} (\coqident{time.Hinnant}{offset_rd}).

\subsubsection{Proofs}
\label{sec:proofs}

\begin{figure*}
  \centering
  \begin{tikzcd}
    \texttt{calendar}.\coqident{time.calendar}{timestamp}
      &&&
      \Hinnant{timestamp}
      \arrow[llldd, "\coqident{time.Hinnant}{cal_from_timestampK}" description]\\
    &&&\\
    \texttt{calendar}.\coqident{time.calendar}{from_timestamp}
      \arrow[uu, "\texttt{calendar}.\coqident{time.calendar}{timestampK}"]
      &&&
      \Hinnant{from_timestamp}
      \arrow[uu, "\Hinnant{timestampK}"']           
  \end{tikzcd}
  \caption{The implementation and specification of the main time functions together with the canceling lemmas used to prove their correctness. Each arrow from $f$ to $g$ represents the proof that $f$ is a left inverse of $g$.}
  \label{fig:proofs}
\end{figure*}
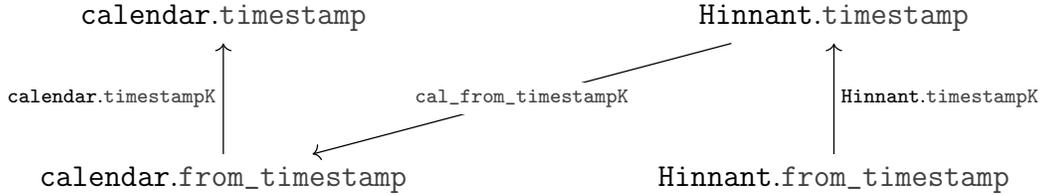

The specification and implementation of the main functions differ significantly, and so it is hard to directly prove that they match. Instead, we make use of lemmas showing that certain functions are the (left) inverse of others (also known as canceling lemmas) and the following simple result.
\begin{remark}
\label{rem:equality_from_canceling}
  Let $T$ and $U$ be types, and $f_1, f_2 : T \to U$ be functions. If there is a function $g : U \to T$ that is both a right inverse of $f_1$ and a left inverse of $f_2$, then $f_1$ and $f_2$ are extensionally equal.
\end{remark}

We summarize the correctness proof for \coqident{time.Hinnant}{timestamp} (Theorem~\ref{thm:timestampE}) as an example. The actual Coq statement includes our standard assumptions on the shape of the list of leap seconds (see Section~\ref{sec:datatypes}), omitted here. Note that the theorem statement is about valid \coqident{time.calendar}{time}s; we make no claim about non-existing times such as any moment during January 32nd. Since \coqident{time.Hinnant}{timestamp} takes a \coqident{time.calendar}{rawTime} as an argument, the implicit coercion \coqident{time.calendar}{rawTime_of_time} (see Figure~\ref{fig:diamond}) is automatically inserted on the left-hand side of the equation. Links to the formalizations of the other proofs can be found in Table~\ref{tab:main_functions:correctness}, which summarizes the results. Note that while the pre-conditions are explicit, the post-conditions are implied by the equality to the specifications.

\begin{theorem}[\coqident{time.Hinnant}{timestampE}]
\label{thm:timestampE}
  For every \coqident{time.calendar}{time} $t$:
    \begin{equation*}
      \Hinnant{timestamp} \ t
      =
      \calendar{timestamp} \ t
      .
    \end{equation*}
\end{theorem}
\begin{proof}
  By Remark~\ref{rem:equality_from_canceling} it suffices to find a suitable function bridging the implementation and specification of timestamp. We used both \calendar{from_timestamp} and the canceling lemmas \calendar{timestampK} and \coqident{time.Hinnant}{cal_from_timestampK} (see Figure~\ref{fig:proofs} for a schematic representation of their statements).
\end{proof}

Given the above strategy, the main challenge becomes proving the canceling lemmas. There are three relevant lemmas for time, depicted as the arrows in Figure~\ref{fig:proofs}, and three analogous ones for dates, used as stepping stones for the time ones and not shown here. We briefly comment on each of these three results.

On the specification side, the \calendar{timestampK} lemma states that \calendar{from_timestamp} is a left inverse of \calendar{timestamp}. Since this is a statement about two specifications, designed to behave nicely with respect to proofs, there were no great difficulties in proving it.

The bridge between the specification and the implementation is provided by the \coqident{time.Hinnant}{cal_from_timestampK} lemma, which states that \calendar{from_timestamp} is a right inverse of \Hinnant{timestamp}. Fortunately, \calendar{from_timestamp} is very simple (just iterating \coqident{time.calendar}{next_time}), and so this proof follows without too much difficulty using basic arithmetical facts.

Finally, on the implementation side, the \Hinnant{timestampK} lemma (needed for the proof of \Hinnant{from_timestampE}, which is the correctness theorem for \Hinnant{from_timestamp}) states that the function \Hinnant{from_timestamp} is a left inverse of \Hinnant{timestamp}. Its proof is the most intricate of the three if done with pen and paper, due to the ubiquitous presence of Euclidean division, which doesn't have an inverse. However, once we developed the automation tool FV Check Range (see Section~\ref{sec:plungin}), the proof was notably eased.

\subsection{Time Arithmetic}
\label{sec:formal}

When adding and subtracting durations to a given time, the irregular periods that the Gregorian calendar and UTC define must be taken into account. For systems that work in Unix, the issue arises with months and years, because they don't have a constant duration. What some systems do is define arithmetical operations on months and years that don't respect basic arithmetical properties \cite{microsoftdatetime}.
  For example, it's common to define the notion of adding 1 month as adding 1 to the month component of the time. However, the result of this operation is not always valid. Thus, adding 1 month in this sense to 2009-01-31 14:00:00 yields 2009-02-31 14:00:00, which is not a valid time because February doesn't have 31 days. The adopted solution is to correct the wrong component by going back to the previous valid one, so the result would be 2009-02-28 14:00:00.
  Similarly, it is possible to add any number to the month component, carrying to the year if necessary, and then correct the wrong component. For example, adding 24 months to 2008-02-29 15:00:00 gives 2010-02-28 15:00:00.
This operation and the analogously defined subtraction are not mutual inverses, since subtracting 1 month from 2009-02-28 14:00:00 would lead to 2009-01-28 14:00:00, which is some days apart from the original 2009-01-31 14:00:00. These are used for practical purposes such as accounting, monthly interest calculation, or utility bills. However, for general computations on durations this behavior may be undesired.

Our library uses UTC, which means that this problem affects all the components except seconds. Not all minutes have the same duration, nor all hours, nor all days. Our solution is to implement two different types of operations in time arithmetic. The first approach, leading to the so-called shift functions, follows the above logic. The second one is a definition of a standard for durations called formal time, and operations called \coqident{time.formalTime}{add_formal} that manipulate fixed amounts of seconds and thus do not suffer from the above issue. Since both options are available, users are free to choose the best one for them.

\subsubsection{Shift Functions}

The shift functions are defined according to the logic described above. Thus, the shift function shifts a component of the time, carrying to the left if necessary, and then if the result is invalid it performs corrections on the wrong component(s) to return a certain close valid time. The precise specification can be found in the lemma \coqident{time.formalTime}{shift_utc_yearsP} for years, and similarly for the other date-time components.


\subsubsection{Formal Time Arithmetic Functions}

In order to have time arithmetic with the usual arithmetical properties, we have defined formal time, which establishes standard durations for every component. A formal second is an atomic second, a formal minute is 60 formal seconds, and so on.

We have chosen to define a formal month as 30 formal days. Therefore, \coqident{time.formalTime}{add_formal_month} adds a constant number of seconds ($30 \cdot 24 \cdot 60 \cdot 60$ seconds) to the input. In the example above, adding a formal month to 2009-01-31 14:00:00 yields 2009-03-02 14:00:00. The result is always valid by construction, except when it goes beyond the minimum or maximum dates. Subtraction works similarly. 

\subsection{Type Refinements}
\label{sec:refinements}

The default representation of the natural numbers in Coq (and MathComp) is the unary one, which uses roughly a symbol per unit.
It is extremely useful for proving properties about the natural numbers, as one can reason by structural induction: prove first that a property holds for zero, and then that if it holds for $n$ it also holds for its successor. However, it is a very inefficient representation in terms of space. For example, explicitly representing numbers larger than $5000$ in Coq makes the code almost unusable. There are ways around this, namely to encode the natural numbers in binary, which is the standard in computer science. The downside is that proofs using natural numbers become more complicated.

The usual solution for this problem is to use what is known as a type refinement. This technique includes an intermediate step where algorithms are defined on top of non-efficient data types and then refined (i.e., redefined) on top of efficient ones. A proof can then be provided to ensure that the refinement kept the relevant properties of the algorithm. This approach has been known and used for quite some time \cite{GeneralRefinements1981}. 

Our particular approach is the following. For a hypothetical function $f$ with input a natural number and output a natural number, we have a specification using the unary \nat. Then we write a first implementation, which would be an efficient algorithm for $f$ except it also uses \nat. We prove a theorem stating that this first implementation behaves as the specification says. The previous steps are described in Section~\ref{sec:main_functions}.

Then we provide a refinement. We define a second implementation of exactly the same algorithm, but this time using the unsigned primitive integers\footnote{Primitive integers, or machine integers, are the integers directly supported by the processor and used physically in memory, with binary representation. Programming languages usually provide a type representing primitive integers.} \Uint{}\codeword{.int} and their operations, and taking the possibility of overflow into account.\footnote{It is necessary to consider overflow because primitive integers are defined cyclically, so that $2^{63}$ is the same as $0$, and so on.} Now we can prove another lemma stating that, whenever the input is small enough, the outputs of both versions coincide. It then follows that the latter implementation meets the specification for such inputs. Analogously, we refine the unary type \ssrint{}.\codeword{int} into \Sint{}\codeword{.int}.

The choice of \Uint{}\codeword{.int} and \Sint{}\codeword{.int} was based both on their efficiency and their good properties with respect to extraction, as described in Section \ref{sec:extraction}.

Even if at first sight the refinement phase may appear much easier than the previous one, in truth the difficulty in our case was comparable and arguably higher, for two different reasons. The first one is that the process of refining functions requires having a good set of rewriting lemmas between the operations of the source and target types, which in our case didn't exist. Hence, this project led to the development of such a set of lemmas, called FV Prim63 to MathComp (see Section~\ref{sec:utils}), which is now available for any future projects. The other reason is the nature of the task itself: ensuring the equivalence between the original and the refined versions requires ensuring that all of the intermediate steps will not overflow, or otherwise finding the appropriate bounds for which they don't. This was in itself a very extensive and quite tedious task, eased by our automation tool FV Check Range (see Section~\ref{sec:plungin}).

\section{FV Prim63 to MathComp}
\label{sec:utils}

FV Prim63 to MathComp provides locked and unlocked conversions between the proof-oriented libraries \ssrnat{} and \ssrint{}, and between the computation-oriented libraries \Uint{} and \Sint{}. It also provides an extensive set of lemmas for rewriting between their respective arithmetical operations, with bounds on the numbers as side conditions when needed.

This tool is independent of FV Time, and can be installed and used from any development that decides to take the same refinement path. As explained in Section \ref{sec:refinements}, our motivation was to link an abstract specification with efficient code for extraction.

A locked version of a function $f$ is a provably equal but not convertible version of $f$. In Coq, a locked version of $f$ is achieved by hiding the body of $f$ behind an opaque dummy constant (see the documentation of \codeword{locked} at \cite{ssreflect_8.17.1}). This prevents the simplification mechanism commonly triggered during proof development from computing the actual value of the function when applied to a specific argument.
The need for locked conversions in our setting comes from certain use cases where large constants are needed on the specification side, and as such are expected to be represented as a \nat{}. However, we have seen that expressing large numbers with \nat{} takes unfeasible amounts of memory. Our solution is to represent such large constants as primitive integers and rely on a coercion \coqident{prim63_mathcomp.ssrnat_Uint63}{nat_of_uint}, i.e., on an automatically inserted translation from primitive unsigned integers to the unary representation of natural numbers. Crucially, this coercion needs to be locked to prevent simplification from computing the unary representation of the number.

\section{FV Check Range}
\label{sec:plungin}

We developed a set of tactics to automatically solve provable decidable goals with up to three free primitive integer variables bounded by a specific primitive integer range. Given a provable goal with base statement of type \bool{} (and thus decidable), and given at most three primitive integer variables and the bounds on which to check them, the tactics identify the desired boolean statement, generate a list with all the primitive integers in the relevant range and use \codeword{vm_compute} \cite{gregoire2002} to confirm that the boolean statement indeed holds for every number in range.

These tactics work rather fast, checking ranges with sizes on the order of $10^5$ in hundredths of seconds and on the order of $10^7$ in two or three seconds (showing an expected linear progression) in one of our machines.

We used these tactics at several points during the development of FV Time and describe here only a particular example:
\begin{align}
  \label{eq:hell}
  \begin{split}
    &\CustomForall{(0 \leq x < 146097)}\\
    &\hspace{5mm} x \geq \frac{h(x)}{365} \cdot 365 + \frac{h(x)}{365 \cdot 4} - \frac{h(x)}{365 \cdot 100} 
  \end{split}
\end{align}
where:
\[h(x) = x - \frac{x}{1460} + \frac{x}{36524} - \frac{x}{146096}. \]
Note that these are natural numbers, and so the division operation is Euclidean division, meaning that, for example, it is not always the case that $\frac{x}{y} \cdot y = x$.

As expected, we found that using our automation was significantly easier and faster than translating pen and paper proofs when proving \eqref{eq:hell}. In particular, the proof inspired by a pen and paper strategy had 400 lines in Coq and took some minutes to compile, while the proof using FV Check Range is a one-liner and takes hundredths of seconds.

There are many other automation tactics available in the Coq ecosystem, some of which can be used to solve goals similar to \eqref{eq:hell} some of the time. See Section~\ref{sec:relatedwork} for a discussion.

\section{Extraction}
\label{sec:extraction}

The concept of extraction is simple: (automatically) translate statements written in Coq to statements written in some other, faster, language \cite{Letouzey2003, LetouzeyPhD, Letouzey2008}. The extraction algorithms are not themselves fully formalized yet (although this formalization is work in progress in the \Metacoq{} project \cite{Sozeau2023}), and so it is possible that errors in the extraction process lead to unexpected discrepancies between the original and the extracted code. Below we describe a method for extracting Coq programs to OCaml so that the resulting OCaml code is clean, reasonably short, and readable.

\subsection{Clean Extraction}

The main idea is to only extract Coq code that already looks as close as possible to OCaml code, so that the extraction plugin has almost nothing to do. The work of translating the arbitrarily complex original Coq code to Coq code representable in OCaml then falls to the programmer instead of the plugin. This extra work has two advantages: first, one obtains control over the extracted OCaml code, and second, one can still reason about the Coq code that originated it. This means that the distance between the verified Coq code and the unverified OCaml code is much smaller than if one simply relied on the extraction plugin without any pre-processing.

When rewriting the original Coq code to make it representable in OCaml, a common issue is translating those partial functions that were defined using dependent types. Consider, as an example of a partial function, Euclidean division on the natural numbers, \codeword{div}. Division is mathematically undefined when the divisor is 0. There are three main ways of implementing such partial functions in Coq, illustrated here with the type signature of division.
\begin{enumerate}
\item Forbid the input:
  \[\codeword{div}_{\textit{gt0}} : \nat \to \codeword{nat_greater_than_0} \to \nat.\]
  \item Output an error: \[\codeword{div}_{\textit{err}} : \nat \to \nat \to \codeword{nat_or_error}.\]
  \item Output a default value: \[\codeword{div}_{\textit{dflt}} : \nat \to \nat \to \nat.\]
\end{enumerate}

Forbidding the input can be done by taking advantage of dependent types, which are not representable in OCaml. Our proposed solution is to rewrite any code that uses dependent types without them, dealing with partiality in some of the other two ways. Outputting an error can be done with an option type, which allows to keep error handling inside Coq, useful if our extracted code is going to work as an external library for other projects. On the other hand, outputting a default value has the advantage of yielding a type signature as simple as possible, which keeps compositionality with other functions. Theorems can then take into account the necessary hypotheses on the input. We also extracted functions in this flavor and used this strategy in the implementation side of our development, as seen in Section~\ref{sec:main_functions}.

This transformation contributes to avoiding the presence of  \codeword{Obj.magic},\footnote{\codeword{Obj.magic} is a low-level OCaml function that allows casting any type to any other type. It is purposefully undocumented because it is not meant for the casual user.} and moreover gives us control about what to do in cases where an undesired input is given, which can happen when executing the extracted code, because dependent types are not expressible in OCaml.

A different problem arises when using Coq libraries that were not purpose-built for extraction, as is the case of MathComp. The pervasive use of canonical structures does not lend itself well to extraction. The mere presence of a boolean equality over an \codeword{eqType} leads to over 20 lines of almost vacuous OCaml code where often the boolean equality for our desired type could be defined in a couple of much more easily understood lines that need to be included anyway.

Here our proposed solution is to avoid MathComp and other external libraries as much as possible when paring down the functions meant for extraction. It has been our experience that most of the benefit of using external libraries is in the wealth of results about the defined functions, and not in the functions themselves. These are usually not that numerous or hard to redefine using only simple Coq features.

Given the above observations, our proposed procedure for clean extraction is as follows:
\begin{enumerate}\setcounter{enumi}{-1} 
  \item Suppose the functions that need to be extracted live in a file \codeword{original.v}.
    \label{step:original}
  \item In a new file (say, \codeword{extraction_file.v}) that does not import
    nor depend on any other file (save perhaps on simple modules such as
    \codeword{List} from the Coq Standard Library), recursively redefine all the
    functions that are meant to be extracted, i.e., redefine the functions in
    \codeword{original.v} as well as every function mentioned in
    \codeword{original.v}, whether defined in the current project or provided by
    MathComp or others. Avoid any Coq features not present in OCaml.
    \label{step:extraction_file}
  \item Extract the functions in \codeword{extraction_file.v}.
    \label{step:extraction}
  \item In another file (say, \codeword{extraction_file_correct.v}) that imports \codeword{original.v}, \codeword{extraction_file.v}, and anything else useful or necessary, show that the extracted functions behave the same as the original ones, possibly under some reasonable assumptions.
    \label{step:correction}
\end{enumerate}

A simple example following this method can be found in \cite{ExtractionExample}.

\subsection{What and How Did We Extract?}

The only part of the Coq code that makes sense to extract are the implemented algorithms.
We started with the functions defined in \coqfile{time}{HinnantR} and \coqfile{time}{formalTimeR} (Step~\ref{step:original}). We then redefined all functions to be extracted together with their dependencies in \coqfile{time}{fvtm_extraction} (Step~\ref{step:extraction_file}) and extracted them in \coqfile{time}{extraction_command} (Step~\ref{step:extraction}). Note that Coq extraction is recursive, so we only needed to list the functions we wished to add to the OCaml user interface, not every function used to define them.
The link between the original functions and the functions to be extracted was provided in \coqfile{time}{fvtm_extraction_correct}.

We used \ExtrOcamlBasic{} and \ExtrOcamlInt{} from Coq's Standard Library to help with the extraction. The former is a small collection of well-accepted translations, such as mapping Coq's \bool{} type to OCaml's, and other such mappings where the types are basically the same in both languages. The latter maps the Coq definitions of \Uint{} and \Sint{} to the very same OCaml module used to implement primitive integers by the Coq kernel.

Our extracted code can be used as a library, taking into account that functions come in two flavors with respect to the way they deal with partiality: we provide versions named \codeword{f_plain} that output default values on ill-formed inputs, corresponding to the $\codeword{f}_{\textit{dflt}}$ described above, and versions named simply \codeword{f} that output a more complex type called \coqident{time.fvtm_extraction}{possibly}, which includes information about problematic inputs and corresponds to $\codeword{f}_{\textit{err}}$. We decided to have the latter because our purpose when extracting was to use the library from other programming languages, for which we wrote a small command-line interface in OCaml that can be compiled as an executable and invoked from any other program, described in Section~\ref{sec:fvtm}. It was thus convenient to extract versions that detect errors (where we proved lemmas about what errors are detected, and when), instead of implementing the full exception handling in OCaml, which would be prone to bugs.

For each function \codeword{f_plain}, there is a lemma \codeword{f_plainR} (where the ``\codeword{R}'' stands for ``refinement'') showing that \codeword{f_plain} meets its specification (i.e., that it behaves like \codeword{calendar.f} on the relevant inputs). Furthermore, there are lemmas proving that the error handling for \codeword{f} is correct given certain assumptions on the input.

\section{FVTM: a Command-line Interface for FV Time}
\label{sec:fvtm}

After extraction, we end up with an OCaml library with all the relevant functions of our development. However, OCaml is not the most popular programming language and communication with other languages is non-trivial. Hence, we wrote a command-line interface in OCaml that allows to compile the library as an executable and invoke it from the terminal or from any other programming language. This command-line version of the library is named FVTM (FV Time Manager) \cite{FVTM_doc}.
The main functions of the library, i.e., the conversions between UTC times and timestamps, can be tried online at \href{https://formalv.com/TimeManager/FVTimeCalculation}{Formal Vindications S.L.'s webpage}.\footnote{\url{https://formalv.com/TimeManager/FVTimeCalculation}}

\section{Related Work}
\label{sec:relatedwork}

The field of formal verification has been flourishing during the last few decades, both in the area of mathematics formalization \cite{Gonthier2008, Gonthier2013_OddOrder, MathComp, ualib2022, Copello2018, ForsterKirst2019, ARQNL2022}, and also in the realm of software verification \cite{FloatingPoint1999, evoting_refinement2007, CompCert2009, ChlipalaCompiler2010, AppelVST2011, Chlipala2011, AppelSHA2562015, FileSystem2015}.

FV Time is not the first library to implement UTC (see for example \cite{googleCPPtime, CPPStdLib, KokaLang, MatLabUTC}), but to the best of our knowledge it is the first formally verified one either for Unix time or for UTC.

Our approach using refinements has been extensively developed both in specific proof assistant developments, such as Coq \cite{Denes2012, Cohen2013, fiat2015} and Isabelle/HOL \cite{IsabelleRefinements2013}. Some other software verification projects have used it too \cite{evoting_refinement2007}.

It must be noted that in the literature many efforts have been devoted to free the developer from the burden of manually proving every result in the proof assistant. These efforts have yielded several Coq tactics to automatically perform certain proof tasks, such as \codeword{micromega} \cite{micromega_tactic}, \codeword{ring} \cite{ring_tactic}, \codeword{interval} \cite{interval}, \codeword{itauto} \cite{itauto_tactic}, \codeword{sauto} \cite{sauto_tactic}, \codeword{firstorder} \cite{firstorder_tactic}, and \codeword{auto} and \codeword{eauto} \cite{auto_tactic_8.17.1}. The tactics \codeword{auto}, \codeword{eauto} and \codeword{sauto} are lemma aggregators, meaning that they produce proofs by combining existing lemmas that can be configured by the user, but this wouldn't solve our intricate arithmetical expressions. The \codeword{ring} tactic is also of no use, since our goals included Euclidean division, which doesn't form a ring, field or semi-field on the integers. As for \codeword{itauto} and \codeword{firstorder}, they are solvers for intuitionistic propositional logic and first order logic respectively, and, although extensible, they can't solve our goals to the best of our attempts.

Notably, \codeword{micromega} provides a set of tactics for arithmetic, one of which can deal with some instances of Euclidean division. In fact, a translation of (\ref{eq:hell}) to \ssrint{} can be automatically solved when using \codeword{mczify} \cite{mczify}, an extension of \codeword{micromega} designed to work with MathComp numbers. However, it can't yet be solved in the realm of binary or primitive integers. This is likely not a fundamental but a practical shortcoming that could be bridged with some work. Nonetheless, our tactics solve any kind of decidable goals on primitive integers, not only arithmetical expressions, and thus the scope is significantly different from \codeword{micromega}'s.

Lastly, \codeword{interval} solves interval arithmetic goals on real numbers. It does not seem like it can solve a translation of our arithmetical expressions in particular. Even if it could, it would introduce a significant amount of overhead and unnecessary axioms due to the detour through the (classical) reals.

\section{Contributions and Conclusion}
\label{sec:concl}

We believe that the development of a formalized time library implementing UTC conversions and operations satisfies an existing need in the panorama of software dealing with time. More importantly, its formal verification led to a number of problems, some of which have yielded general-purpose solutions, while others may be more specific but still inspiring for analogous situations.

The first general-purpose development, FV Prim63 to MathComp, provides a translation between proof-oriented types and operations from MathComp (the \ssrnat{} and the \ssrint{} libraries), and computation-oriented, extraction-friendly types and operations from the Coq Standard Library (the \Uint{} and \Sint{} libraries). We believe that the world of formal verification of software needs powerful, expressive libraries like MathComp for the specification side, and extraction-friendly libraries for the implementation side~\cite{Appel2022}. FV Prim63 to MathComp fills the gap between them.

As direct consequences of this work, it is worth noting that this project catalyzed the addition of signed primitive integers (\Sint{}) to Coq.
The unsigned version was already available, but when we needed the signed version as well, one of the authors teamed up with the Coq developers to make it happen \cite{Sint63}.
Furthermore, we opened a number of minor issues in the Coq bug tracker and helped fix some of them. The improvements live on in the Coq versions since released.

Another general-purpose by-product has been the development of FV Check Range, which provides a set of tactics to automatically solve decidable statements with up to three variables bounded by a specific primitive integer range. These tactics have reduced our development time noticeably, and we hope they serve the same purpose for other teams. Future work includes extending this tool to work with an arbitrary number of variables.

Regarding extraction, we have presented a methodology to minimize possible bugs (crucial before verified extraction for Coq is completed) and obtain clean and simple extracted code. This methodology could serve as a first model of good-practice for other projects.

Finally, we have developed a non-trivial formalized library that is still simple enough to be an accessible example and road map to other libraries. In particular, we have dealt with a number of typical things such as subtyping, partial functions, algorithm and type refinements, extraction, and interface creation.


\section*{Acknowledgments}

  This project wouldn't have been possible without the ideas and help of several people. We are grateful to Guillermo Errezil for prompting the project and to Jasper Hugunin for the idea behind FV Check Range's implementation. We thank Cyril Cohen for his essential help and advice on using Coq. Yannick Forster shared some advice on improving some parts of this paper, for which we are grateful. Finally, we would like to thank the Coq community at large for all the discussions and support. 

  All authors have received research support in a consortium between the University of Barcelona and Formal Vindications S.L.~under the grant RTC-2017-6740-7 of the Spanish Ministry of Science and Universities.


\subsection*{Author Roles}
  \textbf{Ana de Almeida Borges:} Conceptualization, Methodology, Software, Visualization, Writing - original draft, review \& editing;
  \textbf{Mireia González Bedmar:} Conceptualization, Methodology, Software, Visualization, Writing - review \& editing;
  \textbf{Juan Conejero Rodríguez:} Conceptualization, Methodology, Software;
  \textbf{Eduardo Hermo Reyes:} Conceptualization, Software, Writing - review \& editing;
  \textbf{Joaquim Casals Buñuel:} Software, Visualization;
  \textbf{Joost J. Joosten:} Funding acquisition, Project administration, Writing - review \& editing.%
  \footnote{CRediT author statement as described in \url{https://casrai.org/credit/}.}


\balance

\providecommand{\noopsort}[1]{}\providecommand\noopsort[1]{}

\end{document}